%% file: LogSummary.tex
\documentclass[10pt,conference]{IEEEtran}
\IEEEoverridecommandlockouts
\usepackage{cite}
\usepackage{amsmath,amssymb,amsfonts}
\usepackage{algorithmic}
\usepackage{graphicx}
\usepackage{textcomp}
\usepackage{xcolor}

\usepackage[misc]{ifsym} 
\usepackage{times}
\usepackage{multirow}
\usepackage{soul}
\usepackage{url}
\usepackage{graphicx}
\usepackage{amsmath}
\usepackage{amsthm}
\usepackage{booktabs}
\usepackage{algorithm}
\usepackage{algorithmic}
\usepackage{fancyhdr}
\newcommand{\name}{LogSummary}

\def\BibTeX{{\rm B\kern-.05em{\sc i\kern-.025em b}\kern-.08em
    T\kern-.1667em\lower.7ex\hbox{E}\kern-.125emX}}
\begin{document}

\title{Summarizing Unstructured Logs in Online Services}

\author{
      \IEEEauthorblockN{
      Weibin Meng\IEEEauthorrefmark{1},
      Federico Zaiter\IEEEauthorrefmark{1},
      Yuheng Huang\IEEEauthorrefmark{2}, 
      Ying Liu\IEEEauthorrefmark{1},
      Shenglin Zhang
    \IEEEauthorrefmark{3},
      Yuzhe Zhang\IEEEauthorrefmark{3} \\
      Yichen Zhu\IEEEauthorrefmark{4},
      Tianke Zhang \IEEEauthorrefmark{5},
      En Wang\IEEEauthorrefmark{5},
      Zuomin Ren\IEEEauthorrefmark{6},
      Feng Wang\IEEEauthorrefmark{6},
      Shimin Tao\IEEEauthorrefmark{6},
      Dan Pei\IEEEauthorrefmark{1}\\
      }
      \IEEEauthorblockA{
          \IEEEauthorrefmark{1}Tsinghua University, Beijing
National Research Center for Information
Science and Technology\\
          \IEEEauthorrefmark{2}Beijing University of Posts and Telecommunications\\
          \IEEEauthorrefmark{3}Nankai University,
           \IEEEauthorrefmark{4}University of Toronto,
           \IEEEauthorrefmark{5}Jilin University,
            \IEEEauthorrefmark{6}Huawei
      }
  }

\def\ie{\textit{i.e.},~}
\def\etal{\textit{et al.}~}
\def\etc{\textit{etc.}~}
\def\eg{\textit{e.g.},~}
\def\Eg{\textit{E.g.},~}
\def\vs{\textit{vs.}~}
\def\aka{\textit{a.k.a.}~}

\maketitle

\begin{abstract}
\input{abstract}
\end{abstract}


\section{Introduction}
\input{intro.tex}

\label{sec:introduction}

\section{Background}\label{sec:background}
\input{background.tex}

\section{Challenges and Overview}\label{sec:challenge}

\input{challenge.tex}

\section{Algorithm}\label{sec:design}

\input{design.tex}

\section{Experiments}\label{sec:evaluation}
\input{evaluation}

\section{Case Study}\label{sec:case-study}

\input{case-study}

\section{Related Work}\label{sec:related}
\input{related_work}

\section{Conclusion}\label{sec:conclusion}
\input{conclusion}

\bibliographystyle{unsrt} 
\bibliography{references} 

\end{document}

%% file: abstract.tex
Logs are one of the most valuable data sources for managing large-scale online services. 
After a failure is detected/diagnosed/predicted, operators still have to inspect the raw logs to gain a summarized view before take actions.
However, manual or rule-based log summarization has become inefficient and ineffective.
In this work, we propose LogSummary, an automatic, unsupervised end-to-end log summarization framework for online services. 
LogSummary obtains the summarized triples of important logs for a given log sequence.
It integrates a novel information extraction method taking both semantic information and domain knowledge into consideration, with a new triple ranking approach using the global knowledge learned from all logs.
Given the lack of a publicly-available gold standard for log summarization, we have manually labelled the summaries of four open-source log datasets and made them publicly available.
The evaluation on these datasets as well as the case studies on real-world logs demonstrate that LogSummary produces a highly representative (average ROUGE F1 score of 0.741) summaries.
We have packaged LogSummary into an open-source toolkit and hope that it can benefit for future NLP-powered summarization works.


%% file: intro.tex
With the continuous development of Internet applications, large-scale online software service systems (\textit{services} for short hereinafter) are getting increasingly large and complex.
A non-trivial service anomaly can impact the user experience of millions of users and lead to significant revenue loss~\cite{pi2019semantic}.
Consequently, the reliability of online services is of vital importance.

\begin{figure}
      \begin{minipage}{1.0\linewidth}
      \centering
      \includegraphics[width = 8.5 cm]{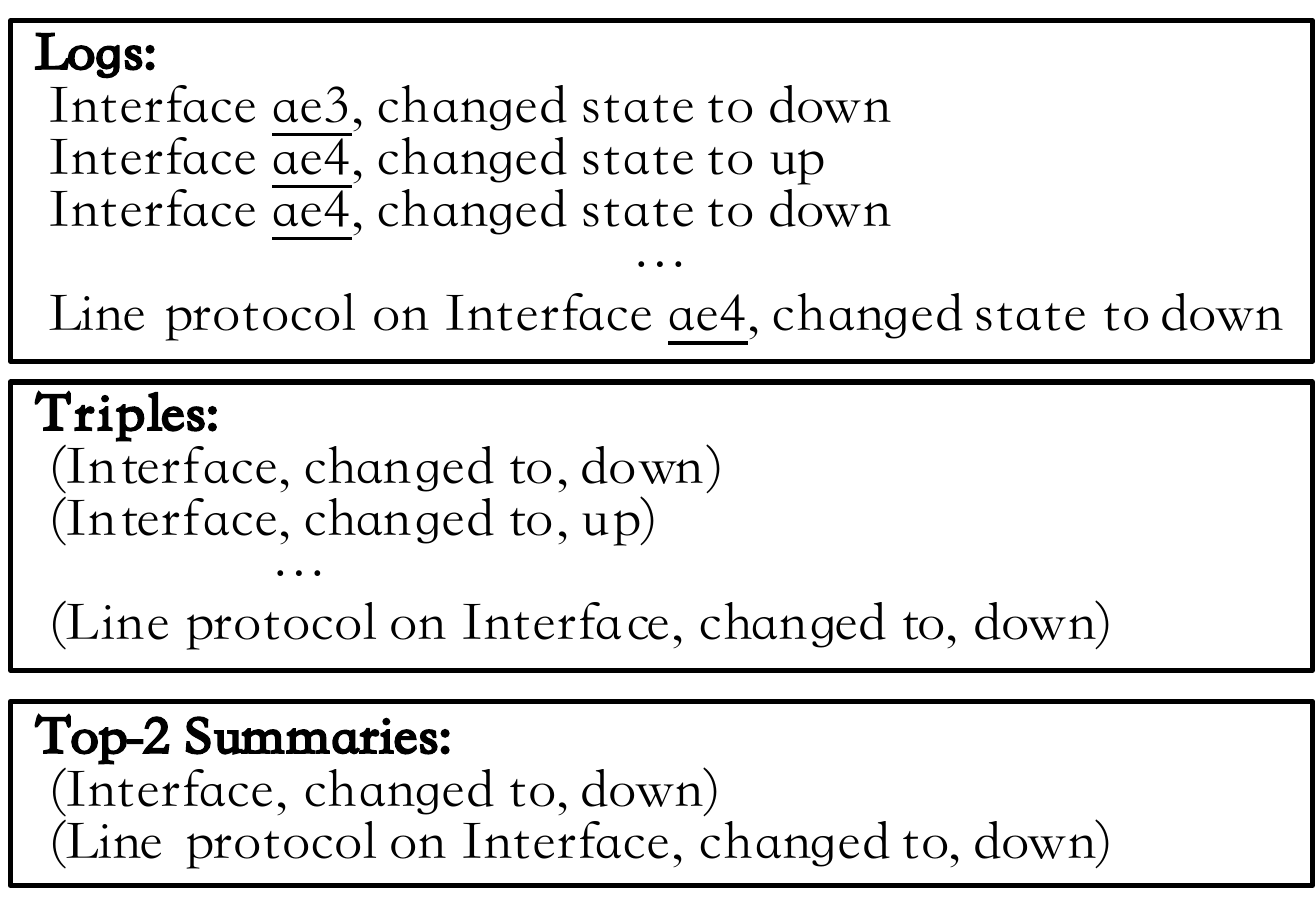}\\
      \end{minipage}
       \caption{Log stream example and summaries.}\label{fig:logs}
\end{figure}

Large-scale services usually generate logs (see the top half of Fig.~\ref{fig:logs}), which describe a vast range of events observed by them and are often the only available data recording service runtime information.
Therefore, many automatic log analysis approaches have been proposed for service management~\cite{he2020survey}, which can be classified into log compression methods (\eg \cite{logzip}), log parsing methods (\eg \cite{logparse,zhu2019tools}), anomaly detection methods (\eg \cite{deeplog,loganomaly}), failure prediction methods (\eg\cite{zhang2018prefix}), failure diagnosis methods (\eg \cite{zhou2019latent}), \emph{etc.}
Although these approaches help operators efficiently understand the status of services, they leave the burden of summarizing logs to operators.
More specifically, after a failure is detected/predicted/diagnosed, operators still have to read the corresponding original logs (\ie a log sequence) to make sense of service-wide semantics~\cite{dogga2019system}.
This is because the existing automatic log analysis approaches, especially those for failure prediction or diagnosis purpose, are not that accurate and general for every scenario. 
Operators must make sure a failure has occurred or will occur before they take measures to mitigate or avoid failures.
 
However, manual log summarization, or the rule (\eg regular expression rule) based log summarization, has become ineffective and inefficient because of the following three reasons.
(1) A large-scale service is usually implemented and maintained by hundreds of developers and operators. 
The developers or operators who investigate logs often have incomplete knowledge of the original logging purpose.
(2) 
The volume of logs is proliferating, for instance, at a rate of about 50 gigabytes (around 120$ \sim $200 million lines) per hour~\cite{zhu2019tools}.
The traditional way of log summarization which largely relies on manual inspection or rule-update has become a labor-intensive and error-prone task.
(3) With agile software development becoming increasingly more popular, operators deploy software updates more frequently, leading to a large number of new types of logs being generated continuously.
It is very difficult for operators to timely comprehend these new types of logs.
Although several works have been proposed for log compression~\cite{gentili2012data, logzip} or system interpretability through logs~\cite{jianguanglou2010mining, ton19}, to the best of our knowledge, there is no previous work aiming to help operators efficiently and effectively summarize a given log sequence in an interpretable and readable manner.
%

When operators investigate logs, they usually care about three key pieces of information, \textit{i.e., } ``entities'',  ``events'' and the ``relation'' between them.
In Fig. \ref{fig:logs}, for example, (``Interface'', ``changed to '', ``down'') is the expected relational triples  extracted from the first log.
These triples are easy to understand for operators because they keep both the semantics and syntax of the original log~\cite{ApplicationsMausam2016OpenIE}.
If we can automatically provide operators the \emph{summarized triples} of the \emph{important} logs in a given log sequence, operators can gain a clear view of this log sequence.
The original goal of service logs, \ie ``\textit{logs are designed for operators to read}'', motivates us to apply natural language processing (NLP) methods to summarize logs.
%
A large number of works have been proposed for text summarization in NLP domain~\cite{jipeng2019short}.
However, there are four following challenges lying in applying existing NLP methods for log summarization.
\begin{enumerate}
\item It is difficult for the existing NLP tools to accurately extract the expected triples from logs,
because logs not only contain normal words, but also domain-specific symbols, and the syntax of a large portion of logs significantly differs from normal sentences.

\item 
Online services can generate a huge number of logs in a short period.
Typically, operators usually investigate all the logs of some period (say one hour before a failure) to obtain the summary of these logs~\cite{gunter2007log}.
Applying existing NLP tools to extract triples for each newly generated log is computationally inefficient (see Table~\ref{table:logie-speed}).

\item  
Existing NLP methods usually summarize texts based on the order of sentences (logs) \cite{kryscinski2019neural}. 
However, for a given log sequence, operators expect to read the summarization of \emph{important} logs first.
Therefore, there is a huge gap between the expectation of operators and the summarization generated by existing NLP methods.

\item 
Applying NLP methods to learn a text summarization model usually needs a large-scale training set~\cite{jipeng2019short}.
However, to the best of our knowledge, although there are publicly available log datasets~\cite{zhu2019tools}, there is no publicly available dataset for log summarization yet.

\end{enumerate}

To address the above challenges, we propose LogSummary, an automatic, unsupervised end-to-end log summarization framework for online services. 
The goal of LogSummary is to obtain the summarized triples of important logs for a given log sequence, which takes both semantic information and domain knowledge into consideration.
The contributions of this paper are summarized as follows.

\begin{enumerate}

\item We propose LogSummary, a framework to perform log summarization for online services.
For a given log sequence, LogSummary obtains the summarized triples of important logs. 
The summary preserves the important information of this log sequence, and is easy to understand.
The implementation of LogSummary is available online\footnote{https://github.com/LogSummary/ICSE2021}.

\item We propose LogIE (Log Information Extraction), a domain-specific and efficient information extraction approach. 
It accurately extracts the expected triples for each log by integrating it with domain knowledge (addressing challenge 1), and achieving efficient information extraction by combining it with log template (addressing challenge 2).

\item 
We propose a simple yet effective method to rank the triples generated by LogIE.
It ranks triples according to the global knowledge learned from all logs, rather than the local information of each triple.
In this way, LogSummary accurately obtains the triples of \emph{important} logs expected by operators (addressing challenge 3).

\item Given the lack of a publicly-available gold standard for log summarization, we manually labelled reference summaries for four existing open-source log datasets and make them available on github\footnote{https://github.com/LogSummary/ICSE2021/tree/main/data}.  We believe that the availability of a summary gold standard would benefit future research and facilitate the adoption of automated log summarization.
\end{enumerate}

The rest of the paper is organized as follows: 
We discuss background in Section \ref{sec:background}, highlight the challenges in Section \ref{sec:challenge} and propose our approach in Section \ref{sec:design}. 
The evaluation is shown in Section \ref{sec:evaluation}.
In Section \ref{sec:related}, we introduce the related works. 
Lastly, we conclude our work in Section \ref{sec:conclusion}.

%% file: background.tex
\textbf{Log Templates.}
Log parsing usually serves as the the first step towards automated log analysis.
The most popular log parsing approach is automatic template extraction~\cite{zhu2019tools,drain,ft-tree,logparse}, which extract constant fields (templates) from logs. 
For example, ``Interface *, changed state to down'' is the template of the first and fifth logs in Fig.~\ref{fig:logs}, and traditional log parsing methods can extract templates from historical logs automatically. 
However, operators continuously conduct software/firmware upgrades on services to introduce new features, fix bugs, or improve performance. These upgrades usually generate new types of logs, and it is required to update templates online~\cite{kabinna2018examining}.
Therefore, we incorporate our previous work, LogParse~\cite{logparse}, to extract templates in this work.  
LogParse~\cite{logparse} is an unsupervised framework which contains a classifier to distinguish the template words and variables at runtime. 
It can extract and learn templates online without retraining its model.



\textbf{Word Embedding for Logs.} 
Logs are designed to facilitate user readability. 
Consequently, the constant parts of logs are defined in human readable manner by developers. 
Many methods (\textit{e.g.,} word2vec~\cite{mikolov2013exploiting}) thus use natural language process (NLP) methods to represent words. 
However, these methods cannot represent domain-specific words accurately. 
For example, ``down'' and ``up'' in Fig.~\ref{fig:logs} are antonyms but they have similar contexts.
Besides, system upgrades usually generate new types of logs with unseen words~\cite{log2vec} (\textit{e.g.,} ``Vlan-interface'' in Figure~\ref{fig:logs}), which pose a challenge for generating distributed representations of words in logs.
For this reason, we adopt our previous work, Log2Vec~\cite{log2vec}, to represent the words of logs.
Log2Vec combines a log-specific word embedding method to accurately extract the semantic information of logs, with an out of vocabulary (OOV) word processor to embed unseen words into distributed representation at runtime.

\textbf{Information Extraction.} Information extraction retrieves relational triples from unstructured text. This is usually done in the form of triples for binary relations, relating two arguments by a predicate or relation for each relation in a given sentence, \textit{i.e.} $(argument_{1}, relation_{1,2},argument_{2})$ \cite{Banko2008OpenIE}. Traditional information extraction approaches rely on predefining a limited set of target relations and hand-crafted patterns. For this they would adapt Named Entity Recognizers and dependency parsers to target a specific domain. These approaches would then require manual effort to be repurposed and applied to a different domain. In order to address the challenge of scalability for performing web information extraction, Banko \textit{et al}. \cite{Banko2008OpenIE} introduced Open Information Extraction (OpenIE). They presented a new information extraction paradigm that would allow to extract relations without defining the number or type of relations in advance.


%% file: challenge.tex

\begin{figure}
      \begin{minipage}[h]{1.0\linewidth}
      \centering
      \includegraphics[width = 8.5 cm]{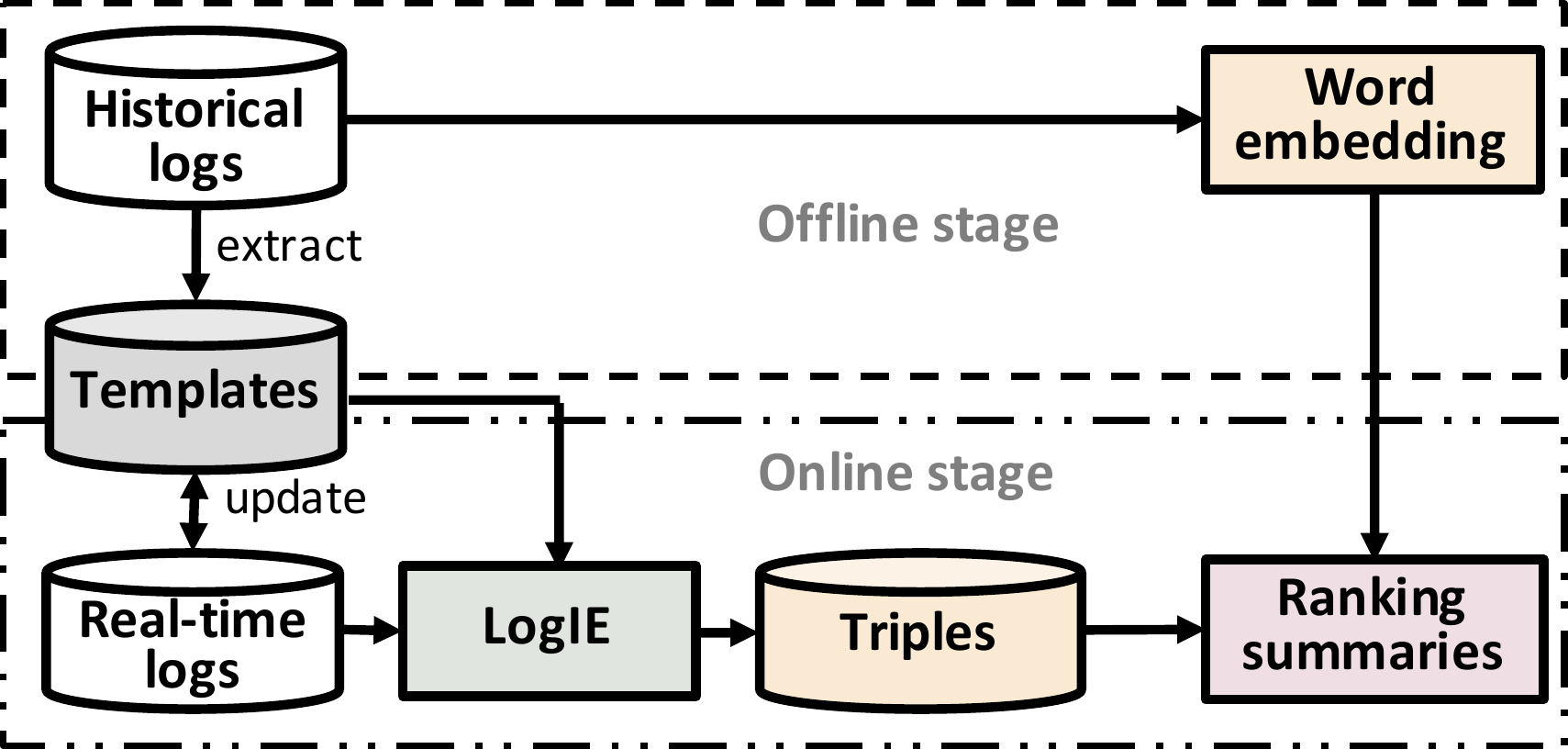}\\
      \end{minipage}
       \caption{Framework of LogSummary}\label{fig:logsummary}
      \vspace{-1 mm}
\end{figure} 

\begin{figure}
      \begin{minipage}[h]{1.0\linewidth}
      \centering
      \includegraphics[width = 8.5 cm]{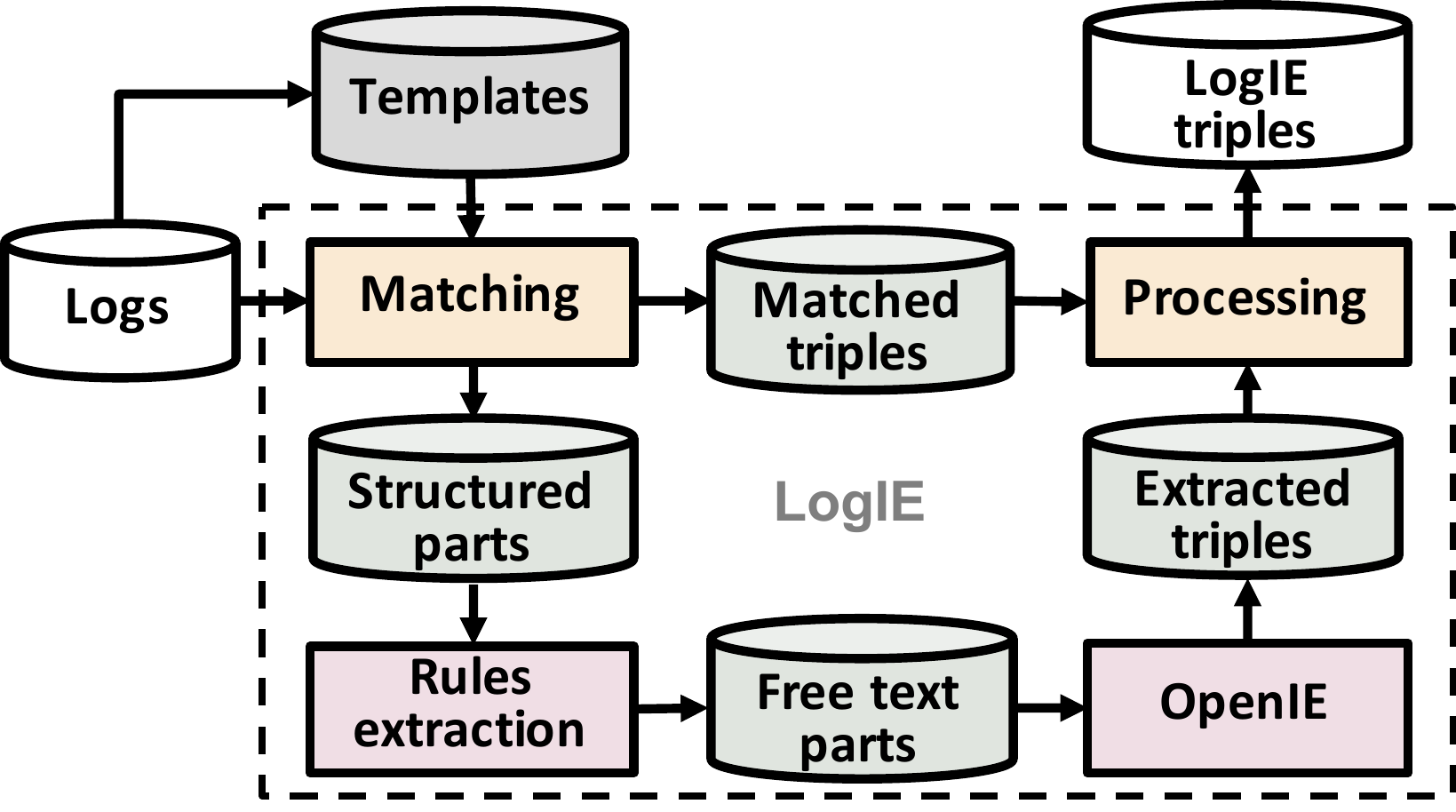}\\
      \end{minipage}
       \caption{Detailed workflow of the LogIE in LogSummary (Fig.~\ref{fig:logsummary}).}\label{fig:logie}
      \vspace{-1 mm}
\end{figure}

The requirement to read and understand logs quickly in online services motivates the design and implementation of LogSummary. 
LogSummary is an automatic log summarization approach which takes both semantic information and domain knowledge into consideration. 

\subsection{Design Challenges}\label{sec:detail-challenge}
Log data is an important data source recording system states and significant events at runtime. 
It is thus intuitive for operators to observe system status and inspect potential anomalous events using logs.
A log is usually printed by logging statements (\eg printf(), logger.info()) in the source code, which are predefined by developers. Typically, the predefined part of a log is human readable. 
Therefore, solving log summarization problems using NLP tools seems promising. 
However, directly applying existing NLP approaches for log summarization faces several challenges as follows.

\noindent\textbf{Domain-specific symbols and grammar.}
Logs contain many domain-specific symbols, and their grammer may significantly differ from normal sentences. 
Existing NLP tools, which are typically designed for normal sentences, cannot get accurate summaries for them. 
For example, entity-value pairs are valuable and structured information that should be extracted from unstructured logs.
However, existing NLP tools cannot extract them directly,
because they may be separated by an equal ``='' or a colon ``:'' symbol. 
Besides, some entity-value pairs are hidden in word combination. 
For instance, when NLP tools process the first log in  Fig. \ref{fig:logs},  it may treat ``Interface ae3'' as a whole, while ``ae3'' is a value for the entity of ``Interface''.

\noindent\textbf{High summarization efficiency requirement.}
After a failure is detected or predicted, operator hope to quickly obtain the summary of a collection of logs in some period (\eg one hour before a detected failure) to figure out what happens on the online service.
However, the online service can generate a large number of logs in this period.
For example, one program execution in the HDFS system generates 288,775 logs per hour~\cite{deeplog}.
On the other hand, existing NLP methods typically get the summarized triples one sentence (log) by one sentence (log), and their efficiency cannot satisfy the requirement of operators (see Table~\ref{table:logie-speed} for more details).

\noindent\textbf{Obtaining the summarized triples of important logs.}
Typically, logs are generated in the order of program executing, and they contain redundancy and repetition.
When operators inspect a collection of logs triggered by a failure detection or prediction, they want to obtain the triples of those \emph{important} logs first.
However, existing NLP approaches usually generate summaries according to the order of sentences (logs) in the original text (log sequence).
In Fig. \ref{fig:case-study-logs}, for example, an state-of-the-art NLP method generates summaries by compressing original logs, instead of generating the triples of the expected important logs.
Consequently, these approaches cannot satisfy the expectation of operators.



\noindent\subsection{Overview of LogSummary}\label{sec:overview}

In this section, we design LogSummary (as shown in Fig.~\ref{fig:logsummary}) to summarize logs in online services and help operators to read/understand logs faster. LogSummary have two parts, offline training and online summarization.

During offline training, LogSummary applies unsupervised template extraction methods \cite{zhu2019tools} to get templates from historical logs. These templates are used by LogIE in the online stage for matching and processing logs.
Since it's nearly impractical to rank arbitrary summaries, LogSummary applies Log2Vec \cite{log2vec} to learn the semantics of logs and train word embeddings. Log2Vec not only learns domain-specific semantics from offline logs, but also generates a new embedding for unseen (OOV) words at runtime, which are common in new generated logs in online systems. Then, LogSummary applies the trained embeddings to rank summaries online.

In the online stage, we design LogIE (describe in Section \ref{sec:logie})  a mechanism to generate triples for given logs. The inputs of LogIE are real-time logs and templates. LogSummary updates new templates automatically by using LogParse \cite{logparse}. LogIE outputs summaries of logs. LogIE solves the challenge that logs contain domain-specific text. Besides, LogIE saves mapping caches between triples and templates, which speeds up log processing and solves the processing speed challenge imposed by the huge amount of logs.
After LogIE, LogSummary ranks triples by adopting TextRank, which meets the requirements of operators reading important summaries first rather than following the program execution logs order.

%% file: design.tex
\subsection{LogIE}\label{sec:logie}
\begin{figure}
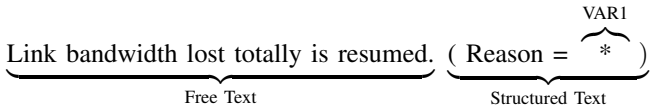

    $\underbrace{\text{Link bandwidth lost totally is resumed.}}_{\text{Free Text}}\ \underbrace{\text{( Reason = }\overbrace{\text{*}}^{\text{VAR1}} \ )}_{\text{Structured Text}}$
    \caption{Log template example.}
    \label{fig:template}
\end{figure}

In order to accurately and efficiently extract valuable information from logs, we propose LogIE (Log Information Extraction). 
LogIE performs open information extraction on logs, extracting triples relating entities and arguments through a relation or predicate. To achieve this, it combines both Rule Extraction (RE) and OpenIE to extract the triples. 
In order to make the process fast and efficient, LogIE adopts templates to improve and speed up information extraction of logs.
Note that, templates are extracted by existing approaches \cite{zhu2019tools,logparse} automatically.
LogIE learns triples from the log templates, so template matching can be used to produce the LogIE triples output.
LogIE is a framework composed of four main components, which we describe and explain how they work using the simple example in Fig. \ref{fig:template}.

\subsubsection{Matching $\And$ Processing}
Matching and Processing are the overarching components of LogIE supported by the RE and the OpenIE components that perform the triple extraction. As shown in Fig. \ref{fig:logie}, LogIE takes both raw logs and templates as its input. LogIE performs template matching on the input rawlogs. Using Fig. \ref{fig:template} as an example, if a log is matched with this template, LogIE retrieves the previously extracted triples for this given log-type and substitute the variables present in the triples by their actual value obtained from the raw log. These variables are usually identifiers, values or service addresses \cite{pi2019semantic}. Therefore once a log is received, if a template is matched by the matching component, it will directly be processed by the Processing component and output its LogIE triples. This way LogIE is able to effectively and efficiently yield OpenIE triples in an online manner. 
Since the goal of LogIE is to get a structured information from logs, we treat all these cases equally by substituting them by a dummy token to be considered an entity or part of an argument \textit{e.g.,} "VARX", where X is the ordinal of the variable within the template.
In the case that the log is not matched to any template, a new template needs to be extracted\cite{drain,logparse}. Since LogIE is meant to be run online, it requires a template extraction and matching method that can be incrementally updated online. For this reason, we incorporate LogParse into the Matching component. The new template is then split into subparts as shown in Fig. \ref{fig:template}, based on rules predefined accordingly to the source services log. These subparts are then handled by the RE and OpenIE components to extract a new set of triples from it. The RE component would handle the structured parts, while the OpenIE one handles the free text parts. The output triples are then stored and passed to the Processing component to produce the final LogIE triples output.

\begin{algorithm}[t] 
\caption{Log Summarization} 
\label{alg:summarization} 
\begin{algorithmic}[1] 
\REQUIRE A semantic information triple set  $ST$, a domain knowledge triple set $DT$,  the number of triple candidates $k$  and word embedding set $WE$ %
\ENSURE Ranked summaries $S$  

\STATE Create a triple vector set $TV$
\FOR{each triples $st$ of $ST$} 
\STATE Create a temporary empty triple vector $tv$
\STATE Let a integer variable $len$ record the number of words in $st$
\FOR{each word $w$ of $st$} 
\STATE Find the corresponding word vector $wv$ for $w$ in $WE$
\STATE Plus the current word vector $wv$ to the temporary triple vector $tv$
\ENDFOR 
\STATE Get the average vector $av$ by dividing $tv$ by $len$ and regard $av$ as the triple vector $st$
\STATE Append the triple vector $av$ to $TV$
\ENDFOR 
\STATE Init a matrix of transition probability $M$ by calculating cosine similarity between all $tv$ pairs in $TV$
\STATE Convert $M$ to a weighted graph $G=(V,E)$. 
\STATE Get the triple scores $TS$ by applying Formula~\ref{equation} to $G$
\STATE Sort the triples in reverse order by scores in $TS$, and the top $k$ triples as the final summaries $S$.


\RETURN $S$
\end{algorithmic} 
\end{algorithm}

\subsubsection{RE for Rule Triples}\label{sec:rule-triples}
The purpose of the RE component is to make the most out of the structure present in logs, namely the structured text part from the example in Fig. \ref{fig:template}.
According to our observation, there are some rules for systems to print logs. Therefore it becomes easier to define rules to extract part of the information present precisely. For example, in our implementation, we use three different rules, where all three cover different ways of representing entity-value pairs. For these cases, entity-value pairs are usually separated by an equals “=” or a colon ``:" symbol. 
Another common case, is formatting such information in the same way command line arguments are specified in a command line interface program. 
The outputs of the RE component of LogIE could be used to provide further details of the log stream in a readable structured manner, or store structured information (entity-value pairs) for further data mining.
Besides, the RE component processes unstructured logs by first extracting triples from the non-free text parts of the logs before the OpenIE component processes their remaining free text parts.

\subsubsection{OpenIE for Semantic Triples}\label{sec:semantic-triples}
Operators pay attention to ``entities'',  ``events'' and the ``relation'' between them when they read logs which making these the most important pieces of information to be considered for log summarization.
OpenIE \cite{Banko2008OpenIE} is usually used to extract relational triples, which is exactly what the operators need, since they are both structured in a human readable way~\cite{ApplicationsMausam2016OpenIE}, and a reduced version of the original logs.
After rule triples were extracted using templates, the remainder free text is passed on to OpenIE component. 
There has been substantial progress on OpenIE approaches since it was proposed by Banko et al. \cite{Banko2008OpenIE}. 
These methods take free text as input and yield OpenIE triples as the output, formed by two arguments related by a predicate e.g. (``Link bandwidth", ``is", ``resumed"). OpenIE methods achieve their objective by leveraging the underlying semantic structure of the sentences for a given language enabling them to find the arguments present and the predicates that relate them. We therefore leverage existing OpenIE approaches in our implementation to fulfill the OpenIE component requirement of the LogIE framework. 
Since LogIE is a framework, none of its components, including OpenIE, are tied to any implementation in particular. 
Besides, many short logs do not have whole three element of triples, \eg, do not contain entity, OpenIE can also generate ``triples'' with less than three elements.
The output of the OpenIE component are the semantic triples, LogSummary will rank them and generate the summary later.
As you will see in our evaluation in Section \ref{sec:evaluation}, we incorporate the main OpenIE methods from the literature into our work as both baselines for LogIE and as part of the LogIE framework for evaluation.

\subsection{Ranking Summaries}\label{sec:summarization}
As aforementioned, logs, which record the status of services in real-time, usually suffer from redundancy and repetition. Traditionally, operators need to read raw logs and extract valuable information manually. 
However, it‘s labor-intensive and time-consuming.
The goal of the LogSummary is to help operators to read/understand logs faster. 
After the LogIE stage, we obtain triples, the minimum units of semantic information within logs. 
In this section, we introduce a mechanism to rank the triples based on their informativeness. 
Operators generally hope to find out the importance of each triple by measuring the connection between each triple and other triples. For a set of logs, most algorithms ignore the semantics and other elements of its words, and simply treat a triple as a collection of words. And each word appears independently and does not depend on each other. However, in log analysis domain, different word combinations have different meanings. 
Operators usually use knowledge drawn from entire logs to make local ranking decisions.
Therefore, we integrate the information of the global corpus into the sorting algorithm in the form of word vectors, and through the combination of the sorting algorithm and word vectors, iteratively score sentences and sort them according to the score.

\subsubsection{Triple Representation}\label{sec:triple-representation}

Firstly, we propose a method to represent triples with domain-specific semantic.
Log2Vec~\cite{log2vec} enables generalization to domain-specific words, which is achieved by integrating the embedding of lexical and relation features into a low-dimensional Euclidean space. 
By training a model over the existing vocabulary, Log2Vec~\cite{log2vec} can later use that model to predict the embedding of any words, even previously unseen words at runtime. 
Therefore, we apply the technique in Log2Vec to represent triples generated from logs.
Leveraging its previous components, we converts any word in the triples into a word embedding vector and generates the triple's vector, which is the weighted average of its word vectors.

\subsubsection{Ranking Triples}\label{sec:triple-rank}


In this section, we propose a method to rank log summaries. 
Its workflow is shown in Algorithm~\ref{alg:summarization}. 
Firstly, we build a graph associated with the logs, where the graph vertices are representative for the units to be ranked. For the application of triple extraction, the goal is to rank entire semantic triples, and therefore a vertex is added to the graph for each triple in logs. 
Same as sentence extraction, we define a relation that determines a connection between two triples if there is a ``similarity'' relation between them, where the ``similarity'' is measured as the cosine similarity~\cite{gravano2003text} of two triples. Note that, other similarity measures (\textit{e.g.,} Euclidean distance~\cite{danielsson1980euclidean}) are also possible.
Such a relation between two triples can be seen as a process of ``recommendation'': given a triple that addresses certain concepts in logs, it is ``recommended'' to refer to other triples in the logs that address similar concepts. 
Therefore a similarity link is drawn between any two triples.

\begin{figure}
      \begin{minipage}[h]{1.0\linewidth}
      \centering
      \includegraphics[width = 8 cm]{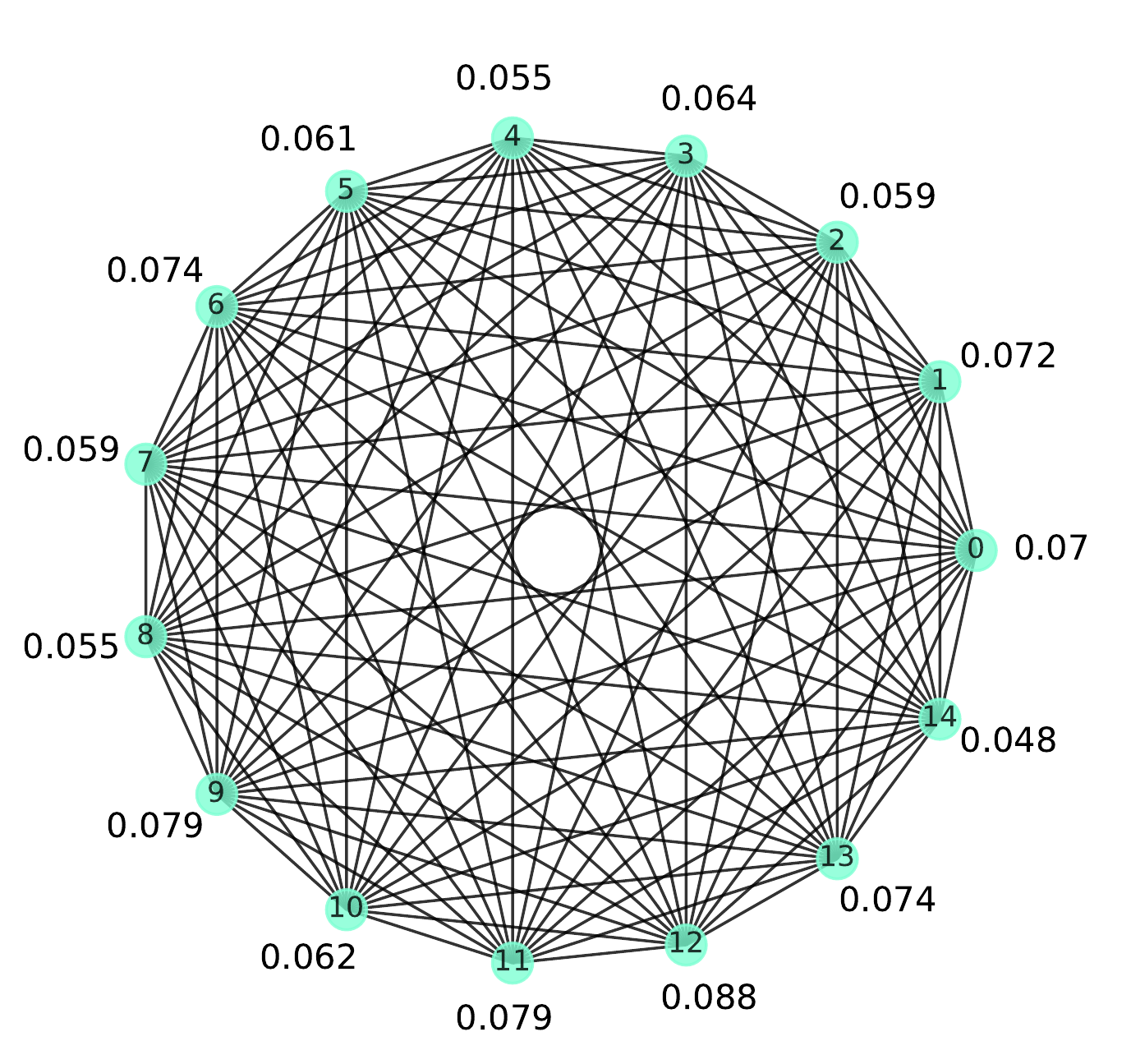}\\
      \end{minipage}

       \caption{Weighted graph of case study on real-world switch logs.}\label{fig:case-graph}
\end{figure} 

Unlike the unweighted graphs in PageRank~\cite{page1999pagerank}, we need to build weighted graphs.
The resulting graph is highly connected, with a weight associated with each edge, indicating the strength of connections established between various triple pairs in logs. The logs are therefore represented as a weighted graph (Fig.~\ref{fig:case-graph} shows a weighted graph for the case study in Section \ref{sec:experiment-summary}).
Formally, let $G=(V, E)$ be a directed graph with the set of vertices $V$ and set of edges $E$, where $E$ is a subset of $V * V$. For a given vertex $V_i$, let $In(V_i)$ be the set of vertices that point to it and let $Out(V_i)$ be the set of vertices that vertex $V_i$ points to.
Then, we adopt the formula in TextRank~\cite{textrank}, which is for graph-based ranking that takes into account edge weights when computing the score associated with a vertex in the graph. Textrank's formula is defined to integrate vertex weights.

\begin{equation}
\label{equation}
WS(V_i) = (1-d) + d*\sum_{V_j\in In_{V_i}}\frac{w_{ji}}{\sum_{V_j\in In_{V_i}}w_{jk}} WS(V_j)
\end{equation}
where $d$ is a damping factor that can be set between 0 and 1.


After the triple-based TextRank~\cite{textrank} is run on the graph, semantic triples are sorted in reverse order of their score.
Note that, although the summaries of LogSummary are highly compressed, it have different goal with other log compression applications (\textit{e.g.,} LogZip\cite{logzip}). 
Other log compression applications aim to store logs while our summary are more readable for operators.


%% file: evaluation.tex
In this section, we report all experiments conducted to evaluate the effectiveness of the proposed LogSummary.
We evaluate LogSummary from two aspects. 
Firstly, we evaluate the accuracy of LogIE on information extraction and compare it with common information extraction approaches. 
Next, we evaluate the accuracy of ranked summaries on the gold references and compare it with the baselines.

\subsection{Experimental Setting}\label{sec:setting}

\subsubsection{Datasets}\label{sec:datasets}
We conduct experiments over four public log datasets from several services: BGL logs~\cite{Xu2009Detecting}, HDFS logs~\cite{deeplog}, HPC logs~\cite{lin2016log}, and Proxifier logs\cite{logzip}.

Since the lack of a publicly-available summary gold standard hinders the automatic evaluation, we manually labelled the above public log datasets and make them available\footnote{https://github.com/LogSummary/ICSE2021/tree/main/data}. 
In this paper, we provide two kinds of gold standard datasets. For information extraction, we labelled all templates for all logs (\cite{zhu2019tools} only label templates for 2000 logs per dataset) and labelled OpenIE triples leveraging semantic information and domain knowledge. 
To evaluate log summarization, we chose 100 groups of 20 contiguous logs per dataset and generated their summaries manually.

The detailed information of above datasets are listed in Table~\ref{tab:datasets}.

\begin{table}[!ht] 
\centering
\caption{Detail of the service log datasets.}\label{tab:datasets}
\renewcommand\tabcolsep{2.75 pt} 
\begin{tabular}{ccc}
\toprule
Datasets&Description&\# of logs\\ 
\midrule
BGL&Blue Gene/L supercomputer log& 4,747,963\\ 
HDFS&Hadoop distributed file system log& 11,175,629 \\
HPC&High performance cluster log& 433,489 \\
Proxifier&Proxifier software log& 21,329 \\

\toprule
\end{tabular}

\end{table}

\subsubsection{Experimental Setup}\label{sec:setup}
We conduct all experiments on a Linux server with Intel Xeon 2.40 GHz CPU.
We implement LogSummary with Python 3.6 and make it open-source\footnote{https://github.com/LogSummary/ICSE2021}.

\subsection{Evaluation on LogIE}\label{sec:experiment-LogIE}
We intrinsically evaluate the LogIE framework in order to choose its best implementation and compare it to the baselines. We evaluate the main OpenIE methods from the literature as baselines and incorporate them as the OpenIE component of LogIE. For this task, we build and open-source a dataset of log information extraction triples based on public logs. We also use this gold standard dataset to evaluate an improved version of LogIE which uses manually improved templates instead of an online template extraction approach. This serves as an ablation test showing the influence of the template quality on LogIE's performance. 

\subsubsection{Triples Gold Standard Dataset}\label{sec:gt-logie}
\begin{table}[t]
\centering
\renewcommand\tabcolsep{8 pt}
\caption{Resulting number of manually improved templates and manually extracted OpenIE triples from the source datasets for building the OpenIE Gold Standard dataset.}
	\label{table:triples-gt-stats}
\begin{tabular}{ccc}
\toprule
    Source &  Templates &  Triples \\
\midrule
       BGL &       263 &         831 \\
      HDFS &        50 &          87 \\
       HPC &        85 &         199 \\
 Proxifier &        14 &          36 \\
    Switch &        95 &         323 \\
\midrule
    Total &        507 &         1476 \\
\bottomrule
\end{tabular}
\end{table}

This dataset, we employ it as the gold standard for building and evaluating the different approaches within LogIE. The process of building it can be divided into three parts: obtaining the logs data from different services; extracting and improving templates used to assist the triples extraction; and manual annotation. 

We sourced the logs from different types of systems and services. Four of them described in Section \ref{table:summary} are open-source and one of them comes from real world switch logs. 

As part of the process of building the gold propositions dataset we extracted templates from all source logs using LogParse \cite{logparse}.

Then, we manually extracted OpenIE triples from the logs. For each log we manually extracted relational triples of the form ($arg_1$, $r$, $arg_2$), meaning that $arg_1$ is related to $arg_2$ by predicate $r$. We aimed to keep the form (subject, predicate, object) in this order. For this purpose considered both the domain knowledge and the semantic structure. We made several considerations:
We extract (subject, predicate, object) triples in this order.  Where applicable, we make prepositions part of the predicate.
It's required to have at least the predicate and the subject or object present to extract a triple.
We make linking verbs the predicate of a triple where applicable.
Conjunctions, such as “and” or “or”, are split into several triples or combined into a single one where the conjunction is part of the predicate.
For the case of adjuncts, regardless of the kind, we decided to disregard them and simply consider them as separators of two different clauses.
For the cases of apposition, we defined an “is” relation, that would serve as an “is-a” relation which is usually used in ontology building. 

Lastly, we leveraged domain knowledge to extract the values of arguments or attributes as well the instances of different entities. 
Usually these would show up as an “=” or a “:” in the logs. In these cases, we also considered an “is” relation to represent the relation between the two. Additionally, arguments are also represented in the format that command line arguments are written. In these cases, we also use an “is” relation and create a second argument “set” for flags.

\subsubsection{Task Formulation}\label{sec:task-logie} As explained in Section \ref{sec:background}, OpenIE intends to obtain all relations present in a given sentence or corpus together with the arguments or entities related by such relations in a structured manner. Likewise, the goal of LogIE is to extract relations present within each log, which are used as the minimum unit of information from each log. Specifically, given a stream of raw logs as the input, relational triples of the form (arg$_1$, relation, arg$_2$) are to be extracted for each relation present within each log. This task will be tested against the gold standard we propose in Section \ref{sec:gt-logie} and evaluated as detailed in the following Section \ref{sec:metric-logie}.

\subsubsection{Metrics and Baselines}\label{sec:metric-logie}
 The main challenge in order to intrinsically evaluate LogIE—similarly to cases that are common in NLP—is that we need to allow different OpenIE triples extractions to be considered acceptable for the same gold proposition. For this reason, we follow a similar approach to that of Stanovsky et al. \cite{Stanovsky2016oieBenchmark} in their OpenIE benchmark. Inspired by He et al. \cite{HeLuheng2015QuestionAnswerDS}, where the syntactic heads of the predicate and the arguments from a given extraction should match those of the corresponding gold proposition, they define a more lenient approach that considers their token-level overlap instead. Therefore, we use an approach similar to theirs\footnote{https://github.com/gabrielStanovsky/oie-benchmark} to calculate precision and recall of the evaluation. Among the main differences, we do not propagate a match to all matching predicates, but thoroughly test all triples against all gold propositions instead. We don't produce a confidence score for each triple, so we don't calculate AUC scores. The main metric we consider for comparison between the approaches is the F1 score. The metrics are calculated as follows:   
$precision = \frac{\#\ correct\ extractions}{\#\ extractions}$, $recall = \frac{\#\ recalled\ gold\ propositions}{\#\ gold\ propositions}$, $F1 = \frac{2\ \times\ precision\ \times\ recall}{precision\ +\ recall}$.

We compare LogIE with six Open Information Extraction methods, namely, ClausIE \cite{Corro2013ClausIECO}, Ollie \cite{Mausam2012OpenLL}, OpenIE5\footnote{https://github.com/dair-iitd/OpenIE-standalone}, PredPatt \cite{Stanovsky2016GettingMO}, PropS \cite{Stanovsky2016GettingMO} and Stanford OpenIE \cite{Angeli2015LeveragingLS}.

\subsubsection{Experimental Results}\label{sec:eval-public-logie}
We evaluate LogIE and compare it against its manually augmented version and the six OpenIE baselines in Table \ref{table:logie} on the four public logs described in Table \ref{tab:datasets}. 
LogIE learns online templates generated by LogParse \cite{logparse}. However, in order to  perform an ablation test, we also compare LogIE Improved which is an augmented version using manually improved templates that were produced as part of the gold standard dataset introduced in Section \ref{sec:gt-logie}. Then each of the baselines are plain OpenIE approaches used directly on the raw logs. LogIE consistently produces better results across all public logs when compared to the baselines. This is because the pipeline approach of LogIE is optimized to make the most of both the structure and the free text present in logs. On the other hand, plain OpenIE methods are actually meant to be used directly on free natural language text. As you will see in Table \ref{table:logie}, even though both versions of LogIE are consistently superior than the baselines, there are cases where their results could be comparable such as on BGL or HPC. The more free text is present in the logs, the easier it is for plain OpenIE methods to generate correct OpenIE triples. However, as we show in Table \ref{table:logie-speed} applying plain OpenIE methods on the logs is not efficient when compared to LogIE which leverages the templates used as input and the high speed of template matching using tries to produce the OpenIE triples output from the raw logs. The throughput of LogIE is over 200X that of applying plain OpenIE. Nonetheless, the performance of LogIE is sensitive to the accuracy of the templates used as input as shown in the ablation test comparison. As demonstrated by the performance of LogIE Improved, the more accurate templates are, the better the performance of LogIE. Further, LogIE leverages either the structure of the log or the semantic information of the unstructured text within logs to extract information. If there is no rich information in the structure, or if details within the log are omitted to make it shorter, its performance is also affected. This is the case for the HDFS logs where the structured parts don't provide rich information and the natural language implicitly refers to the arguments, which is not picked up by LogIE. This affects its results with a low recall as seen in Table \ref{table:logie}. In turn, this affects the output of LogSummary given the pipeline nature of the framework.
\ 
\smallskip

\begin{table}[t]
\centering
\caption{Test accuracy on the log OpenIE triples gold reference dataset of public logs.}
	\label{table:logie}
\begin{tabular}{ccccc}
\toprule
               Logs &                   Method &       Precision &          Recall &              F1 \\
\midrule
       \multirow{7}*{BGL} &  \textbf{LogIE} &  \textbf{0.918} &  \textbf{0.864} &   \textbf{0.89} \\
                    &                  OpenIE5 &           0.788 &           0.733 &           0.760 \\
                    &                 Stanford &           0.685 &           0.753 &           0.717 \\
                    &                    Ollie &           0.552 &           0.633 &           0.590 \\
                    &                 PredPatt &           0.463 &           0.638 &           0.536 \\
                    &                  ClausIE &           0.447 &           0.602 &           0.513 \\
                    &                    PropS &           0.000 &           0.000 &           0.000 \\
      \multirow{7}*{HDFS} &  \textbf{LogIE} &   \textbf{0.98} &  \textbf{0.459} &  \textbf{0.626} \\
                    &                  ClausIE &           0.159 &           0.530 &           0.244 \\
                    &                  OpenIE5 &           0.271 &           0.220 &           0.243 \\
                    &                 Stanford &           0.184 &           0.210 &           0.196 \\
                    &                 PredPatt &           0.171 &           0.177 &           0.174 \\
                    &                    Ollie &           0.003 &           0.079 &           0.006 \\
                    &                    PropS &           0.000 &           0.000 &           0.000 \\
      \multirow{7}*{HPC} &  \textbf{LogIE} &  \textbf{0.859} &  \textbf{0.667} &  \textbf{0.751} \\
                    &                  ClausIE &           0.588 &           0.648 &           0.616 \\
                    &                 PredPatt &           0.591 &           0.556 &           0.573 \\
                    &                 Stanford &           0.691 &           0.349 &           0.464 \\
                    &                    Ollie &           0.290 &           0.285 &           0.287 \\
                    &                  OpenIE5 &           0.567 &           0.123 &           0.202 \\
                    &                    PropS &           0.000 &           0.000 &           0.000 \\
      \multirow{7}*{Proxifier} &  \textbf{LogIE} &  \textbf{0.869} &  \textbf{0.812} &  \textbf{0.839} \\
                    &                 Stanford &           0.831 &           0.254 &           0.389 \\
                    &                  ClausIE &           0.247 &           0.719 &           0.368 \\
                    &                  OpenIE5 &           0.759 &           0.204 &           0.322 \\
                    &                    Ollie &           0.556 &           0.194 &           0.288 \\
                    &                 PredPatt &           0.061 &           0.106 &           0.078 \\
                    &                    PropS &           0.000 &           0.000 &           0.000 \\
\bottomrule
\end{tabular}
\end{table}

\begin{table}[t]
\centering
\renewcommand\tabcolsep{18 pt}
\caption{Comparison of speed measured in logs per second between LogIE and the plain OpenIE methods when processing the input logs measured over thirty runs for each OpenIE method and each logs dataset.}
	\label{table:logie-speed}
\begin{tabular}{ccc}
\toprule
\textbf{Approach} & \multicolumn{2}{l}{\textbf{Throughput (logs / s)}} \\
\cmidrule{2-3}
{} &     mean &      std \\
\midrule
LogIE    & 8550.66 & 1909.62 \\
\midrule
OpenIE Methods &   39.05 &   36.19 \\
\bottomrule
\end{tabular}
\end{table}


\subsection{Evaluation on Ranking Summaries}\label{sec:experiment-summary}

\subsubsection{Metrics and Baselines}\label{sec:metric-summary}
To automatically evaluate the log summarization performance of different approaches, we use ROUGE~\cite{rouge}. The ROUGE metric measures the summary quality by counting the overlapping units between the generated summary and reference summaries. In our scenario, different operators may manually generate summaries in different order of words/phrases. Therefore, we apply ROUGE-1 to evaluate performance.
Following the common practice~\cite{rouge2}, we report the precision, recall, F1 score for ROUGE-1, where $precision = \frac{\#\ overlapping\ words}{\#\ words\ in \ gold reference}$, $recall = \frac{\#\ overlapping\ words}{\#\ words\ in \ automatic \ summary}$ and $F1 socre = \frac{2 \times precision \times recall}{precision + recall}$  We obtain the metrics using open-source package\footnote{https://github.com/pltrdy/rouge}.
We apply the compression ratio, \textit{i.e.,} $\frac{size~of~summaries}{size~of~original~logs}$, to evaluate the log compression performance.
We compare LogSummary with three extractive summarization methods, namely, TF-IDF~\cite{tf-idf}, LDA~\cite{ton19}, TextRank (sentence summary)~\cite{textrank}). 

\subsubsection{Experimental Results}\label{sec:public-summary}
We compare LogSummary with three baselines on four public datasets.
For LogSummary, we choose top-5 semantic triples from online logs.
Table~\ref{table:summary} show the comparison results of \name{} and three baselins. Overall, \name{} achieves the best summarization accuracy among the four methods.
Both TF-IDF and LDA, however, have low F1 scores ($< 0.5$) on all four datasets, because TF-IDF and LDA generate summaries by extracting keywords, which dismisses valuable information in raw logs.
Although TextRank achieves relatively high precision (\textit{e.g.}, 0.904 on the BGL dataset), the high precision is at the cost of low recalls. For instance, on the Proxifier dataset, the recall of TextRank is only 0.050. Because there are many similar logs with different variables, when employed on its own, TextRank may choose many logs of the same type and ignore other types of logs. 
On the contrary, LogSummary, uses LogIE to extract triples from logs as an intermediate representation, which is more fine grained than each complete log, before applying TextRank achieving a $\approx$4.6 times higher recall.

In Table~\ref{table:summary} ,we evaluate the compression ratio for log summarization on four datasets. We find that LogSummary achieves average compression ratio of 3.1\%, which will vastly reduce the reading and understanding load of operators.

The results mean that the outputs of LogSummary are not only readable but also highly compressed.
\begin{table}
\centering
\caption{Log summarization performance and Compression Ratio (CR) of LogSummary compared to its Baselines.}
	\label{table:summary}
\begin{tabular}{cccccc}
\toprule
             Logs &      Method& Precision & Recall & F1 & CR  \\
\midrule

\multirow{4}*{BGL}&  \textbf{LogSummary} &\textbf{0.815} &\textbf{0.703} &\textbf{0.725} &\textbf{0.026} \\
                  &   LDA      &   0.382   &  0.076 &   0.119 & 0.130\\
                  & TextRank   &   0.893   &  0.238 &   0.347 & 0.144\\
                  & TF-IDF    &   0.383   &  0.354 &   0.332 & 0.024 \\
                  
\multirow{4}*{HDFS}&  \textbf{LogSummary} & \textbf{0.759} &\textbf{0.432}& \textbf{0.538} & \textbf{0.015}\\
                  &   LDA      &   0.220   &  0.045 &   0.074 & 0.225\\
                  & TextRank   &   0.602   &  0.079 &   0.135 & 0.230\\
                  & TF-IDF    &   0.193   &  0.176 &   0.179 & 0.033 \\

\multirow{4}*{HPC}&  \textbf{LogSummary}&\textbf{0.819}&\textbf{0.911}&\textbf{0.840} &\textbf{0.037}\\
                  &   LDA      &   0.530   &  0.110 &   0.175 & 0.251 \\
                  & TextRank   &   0.904   &  0.265 &   0.365 & 0.208\\
                  &  TF-IDF    &   0.487   &  0.506 &   0.472 & 0.039 \\
 
\multirow{4}*{Proxifier}&   \textbf{LogSummary} &\textbf{0.879}&\textbf{0.857} &\textbf{0.864} &\textbf{0.045}\\
                  &   LDA      &   0.332   &  0.088 &   0.135 & 0.099\\
                  & TextRank   &   0.663   &  0.050 &   0.093 & 0.275\\
                  & TF-IDF&   0.281   &  0.324 &   0.290 & 0.023\\
                  
\bottomrule
\end{tabular}
\end{table}

\begin{figure}
      \begin{minipage}[h]{1.0\linewidth}
      \centering
      \includegraphics[width = 8.5 cm]{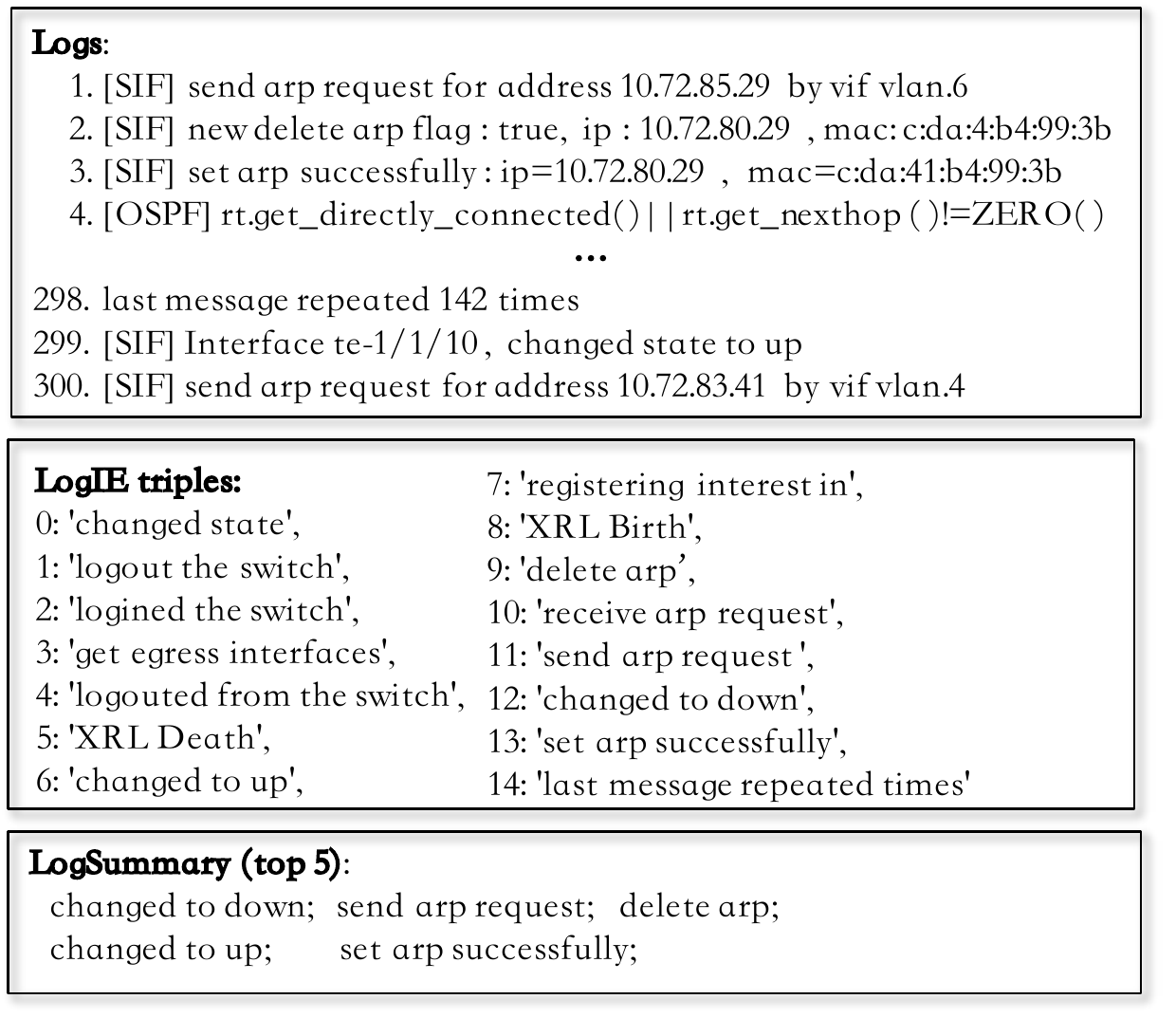}\\
      \end{minipage}
      \vspace{-1 mm}
       \caption{A case study of LogSummary on switch logs.}\label{fig:case-study-logs}
\end{figure}

\subsection{Threats to Validility}
The LogSummary framework leverages each of its components to automatically produce accurate summaries. 
However, its pipeline nature makes each component dependant on the quality of the output from the previous ones. 
Sometimes, we should proceed to improve the templates manually since template extraction is not perfectly precise. 
These imprecisions meant there would be redundant templates extracted from the logs. Additionally, in some cases, the variables may not have been detected properly.
Therefore, the quality of the templates may affect the triples extracted by LogIE, which in turn affects the representations built by Log2Vec \cite{log2vec} which are used by TextRank \cite{textrank} to produce the ranked summaries. 
Nonetheless, each of the components provides significant benefits over their baselines. 
LogIE produces triples at over 200 times the throughput of plain OpenIE methods, which serve as the intermediate result that LogSummary leverages to achieve $\approx$4.6 times the recall of TextRank \cite{textrank}, which is the best performing baseline.
LogSummary can serve further downstream purposes, which we consider for future work. 
The triples of LogSummary could aid the creation of knowledge graphs applied to perform automatic root cause analysis. Additionally they could serve as the intermediate representation before other log analysis tasks.
LogSummary could produce the summary of anomalous logs for troubleshooting.
Further, we consider five log datasets as part of the evaluation from online system services. Our approaches outperform their baselines in all of them, which shows the generalizability of LogSummary. However, it may encounter challenges when dealing with complex application layer logs, \textit{e.g.,} operators create complex rules for the Rule Extraction part.

%% file: case-study.tex
To further evaluate the performance of LogIE, we did a case study on real-world logs, which are generated by switches deployed in a top cloud service provider. In this part, we selected and labelled one million switch logs in the same manner we did for the public datasets as described in Section \ref{sec:gt-logie}. Here results are consistent with the experiments on the public log datasets. Likewise, given that the proportion of free text is higher within this logs dataset, a complex OpenIE approach can also generate comparably accurate triples. Nonetheless, both LogIE approaches outperform all of the baselines at a throughput that is over 200 times higher. 
For this real-world dataset, LogIE achieves F1 of 0.831, where other OpenIE methods get average F1 of 0.414.

Besides, we conducted a case study of LogSummary on switch logs to visualize its intermediate steps and showcase its results.
As shown in Figure~\ref{fig:case-study-logs}, we randomly selected 300 logs from real-world switch logs, applied LogIE to extract triples, and generated a summary with LogSummary. The corresponding weighted triple graph from TextRank is shown in Figure~\ref{fig:case-graph}.
This summarization result was confirmed by operators and prove that LogSummary is useful.

%% file: related_work.tex

Different kinds of log summarization approaches have been proposed for purposes different from ours. 
At the same time, different methods have been used for similar purposes as to enhance the interpretability of the logs for troubleshooting. 

Gentili \textit{et al.} \cite{gentili2012data} proposed an approach to leverage a taxonomy graph over the events presented by the logs in order to reduce the resulting raw data size to keep the required space to store the data manageable and improve the performance and system load for log analysis. 
Gunter \textit{et al.} \cite{gunter2007log} presented an approach for log summarization for the goal of decreasing the load on the system for logging. They do not focus on human interpretability of the logs, but rather on reducing system workload for downstream tasks. Their main evaluation metric is keeping the same performance of autonomous anomaly detection despite the compressed logs.
Shang and Syer focus on understanding software logs \cite{shang2014understanding} and examining the stability of logs \cite{kabinna2018examining}.
He \textit{et al.} \cite{he2018identifying} presented a method to identify impactful service problems by using log analysis. They pay more attention to visualization rather than readable.

Satpathi \textit{et al.} \cite{ton19} proposed the first closely related work in our scenario. Their focus is different from ours as they aim to mine the distribution of messages for each anomalous event and their occurrences in the logs getting an event signature that represents it formed by keywords. Nonetheless, they use an approach that resembles log summarization by first aggregating all the events from the given distribution using their proposed change-point detection and then performing summarization using LDA \cite{Blei2003Latent}. However, LDA has proven not to work properly for short text summarization despite aggregation or previous clustering. Additionally, they do not consider parameters from the logs that are key to understanding an event after summarization and helping operators with their troubleshooting.

%% file: conclusion.tex
Logs play an important role in service maintenance. 
Operators still have to conduct log summarization before take actions in a manual or rule-based manner, even though many methods have been proposed for automatic log detection/diagnosis/prediction.
In this paper, we propose LogSummary, a framework towards automatic summarization for large-scale online services. 
LogSummary combines LogIE, which accurately and efficiently obtains information extraction triples from logs, and a simple yet effective triple ranking method, which utilizes the global knowledge learned from all historical logs.
We perform extensive evaluation experiments to demonstrate LogSummary's performance in summarizing logs. 
Moreover, we have open-sourced LogSummary and the manually labelled gold standard references, hoping that they can benefit for future research works.


%% file: LogSummary.bbl
\begin{thebibliography}{10}

\bibitem{pi2019semantic}
Aidi Pi, Wei Chen, Shaoqi Wang, and Xiaobo Zhou.
\newblock Semantic-aware workflow construction and analysis for distributed
  data analytics systems.
\newblock In {\em Proceedings of the 28th International Symposium on
  High-Performance Parallel and Distributed Computing}, pages 255--266. ACM,
  2019.

\bibitem{he2020survey}
Shilin He, Pinjia He, Zhuangbin Chen, Tianyi Yang, Yuxin Su, and Michael~R Lyu.
\newblock A survey on automated log analysis for reliability engineering.
\newblock {\em arXiv preprint arXiv:2009.07237}, 2020.

\bibitem{logzip}
Jinyang Liu, Jieming Zhu, Shilin He, Pinjia He, Zibin Zheng, and Michael~R Lyu.
\newblock Logzip: Extracting hidden structures via iterative clustering for log
  compression.
\newblock In {\em 2019 34th IEEE/ACM International Conference on Automated
  Software Engineering (ASE)}, pages 863--873. IEEE, 2019.

\bibitem{logparse}
Weibin Meng, Ying Liu, Federico Zaiter, Shenglin Zhang, Yihao Chen, Yuzhe
  Zhang, Yichen Zhu, En~Wang, Ruizhi Zhang, Shimin Tao, Dian Yang, Rong Zhou,
  and Dan Pei.
\newblock Logparse: Making log parsing adaptive through word classification.
\newblock In {\em International Conference on Computer Communication and
  Networks (ICCCN)}, pages 1--9. IEEE, 2020.

\bibitem{zhu2019tools}
Jieming Zhu, Shilin He, Jinyang Liu, Pinjia He, Qi~Xie, Zibin Zheng, and
  Michael~R Lyu.
\newblock Tools and benchmarks for automated log parsing.
\newblock In {\em Proceedings of the 41st International Conference on Software
  Engineering(ICSE)}, pages 121--130, 2019.

\bibitem{deeplog}
Min Du, Feifei Li, Guineng Zheng, and Vivek Srikumar.
\newblock Deeplog: Anomaly detection and diagnosis from system logs through
  deep learning.
\newblock In {\em ACM SIGSAC Conference on Computer and Communications Security
  (CCS)}, pages 1285--1298. ACM, 2017.

\bibitem{loganomaly}
Weibin Meng, Ying Liu, Yichen Zhu, Shenglin Zhang, Dan Pei, Yuqing Liu, Yihao
  Chen, Ruizhi Zhang, Shimin Tao, Pei Sun, et~al.
\newblock Loganomaly: Unsupervised detection of sequential and quantitative
  anomalies in unstructured logs.
\newblock In {\em Proceedings of the Twenty-Eighth International Joint
  Conference on Artificial Intelligence, IJCAI-19. International Joint
  Conferences on Artificial Intelligence Organization}, volume~7, pages
  4739--4745, 2019.

\bibitem{zhang2018prefix}
Shenglin Zhang, Ying Liu, Weibin Meng, Zhiling Luo, Jiahao Bu, Sen Yang,
  Peixian Liang, Dan Pei, Jun Xu, Yuzhi Zhang, et~al.
\newblock Prefix: Switch failure prediction in datacenter networks.
\newblock {\em Proceedings of the ACM on Measurement and Analysis of Computing
  Systems(SIGMETRICS)}, 2(1):2, 2018.

\bibitem{zhou2019latent}
Xiang Zhou, Xin Peng, Tao Xie, Jun Sun, Chao Ji, Dewei Liu, Qilin Xiang, and
  Chuan He.
\newblock Latent error prediction and fault localization for microservice
  applications by learning from system trace logs.
\newblock In {\em Proceedings of the 2019 27th ACM Joint Meeting on European
  Software Engineering Conference and Symposium on the Foundations of Software
  Engineering}, pages 683--694, 2019.

\bibitem{dogga2019system}
Pradeep Dogga, Karthik Narasimhan, Anirudh Sivaraman, and Ravi Netravali.
\newblock A system-wide debugging assistant powered by natural language
  processing.
\newblock In {\em Proceedings of the ACM Symposium on Cloud Computing}, pages
  171--177, 2019.

\bibitem{gentili2012data}
Eleonora Gentili, Alfredo Milani, and Valentina Poggioni.
\newblock Data summarization model for user action log files.
\newblock In {\em International Conference on Computational Science and Its
  Applications}, pages 539--549. Springer, 2012.

\bibitem{jianguanglou2010mining}
Jian-Guang Lou, Qiang Fu, Shengqi Yang, Jiang Li, and Bin Wu.
\newblock Mining program workflow from interleaved traces.
\newblock In {\em Proceedings of the 16th ACM SIGKDD international conference
  on Knowledge discovery and data mining}, pages 613--622. ACM, 2010.

\bibitem{ton19}
Siddhartha Satpathi, Supratim Deb, R~Srikant, and He~Yan.
\newblock Learning latent events from network message logs.
\newblock {\em IEEE/ACM Transactions on Networking}, 27(4):1728--1741, 2019.

\bibitem{ApplicationsMausam2016OpenIE}
Mausam.
\newblock Open information extraction systems and downstream applications.
\newblock In {\em IJCAI}, 2016.

\bibitem{jipeng2019short}
Qiang Jipeng, Qian Zhenyu, Li~Yun, Yuan Yunhao, and Wu~Xindong.
\newblock Short text topic modeling techniques, applications, and performance:
  A survey.
\newblock {\em arXiv preprint arXiv:1904.07695}, 2019.

\bibitem{gunter2007log}
Dan Gunter, Brian~L Tierney, Aaron Brown, Martin Swany, John Bresnahan, and
  Jennifer~M Schopf.
\newblock Log summarization and anomaly detection for troubleshooting
  distributed systems.
\newblock In {\em 2007 8th IEEE/ACM International Conference on Grid
  Computing}, pages 226--234. IEEE, 2007.

\bibitem{kryscinski2019neural}
Wojciech Kry{\'s}ci{\'n}ski, Nitish~Shirish Keskar, Bryan McCann, Caiming
  Xiong, and Richard Socher.
\newblock Neural text summarization: A critical evaluation.
\newblock {\em arXiv preprint arXiv:1908.08960}, 2019.

\bibitem{drain}
Pinjia He, Jieming Zhu, Zibin Zheng, and Michael~R Lyu.
\newblock Drain: An online log parsing approach with fixed depth tree.
\newblock In {\em 2017 IEEE International Conference on Web Services (ICWS)},
  pages 33--40. IEEE, 2017.

\bibitem{ft-tree}
Shenglin Zhang, Weibin Meng, Jiahao Bu, Sen Yang, Ying Liu, Dan Pei, Jun Xu,
  Yu~Chen, Hui Dong, Xianping Qu, et~al.
\newblock Syslog processing for switch failure diagnosis and prediction in
  datacenter networks.
\newblock In {\em 2017 IEEE/ACM 25th International Symposium on Quality of
  Service (IWQoS)}, pages 1--10. IEEE, 2017.

\bibitem{kabinna2018examining}
Suhas Kabinna, Cor-Paul Bezemer, Weiyi Shang, Mark~D Syer, and Ahmed~E Hassan.
\newblock Examining the stability of logging statements.
\newblock {\em Empirical Software Engineering}, 23(1):290--333, 2018.

\bibitem{mikolov2013exploiting}
Tomas Mikolov, Quoc~V Le, and Ilya Sutskever.
\newblock Exploiting similarities among languages for machine translation.
\newblock {\em arXiv:1309.4168}, 2013.

\bibitem{log2vec}
Weibin Meng, Ying Liu, Yuheng Huang, Shenglin Zhang, Federico Zaiter, Bingjin
  Chen, and Dan Pei.
\newblock A semantic-aware representation framework for online log analysis.
\newblock In {\em IEEE International Conference on Computer Communications
  (ICCCN)}, 2020.

\bibitem{Banko2008OpenIE}
Michele Banko, Michael~J. Cafarella, Stephen Soderland, M.~Alexander Broadhead,
  and Oren Etzioni.
\newblock Open information extraction from the web.
\newblock In {\em CACM}, 2008.

\bibitem{gravano2003text}
Luis Gravano, Panagiotis~G Ipeirotis, Nick Koudas, and Divesh Srivastava.
\newblock Text joins for data cleansing and integration in an rdbms.
\newblock In {\em Proceedings 19th International Conference on Data Engineering
  (Cat. No. 03CH37405)}, pages 729--731. IEEE, 2003.

\bibitem{danielsson1980euclidean}
Per-Erik Danielsson.
\newblock Euclidean distance mapping.
\newblock {\em Computer Graphics and image processing}, 14(3):227--248, 1980.

\bibitem{page1999pagerank}
Lawrence Page, Sergey Brin, Rajeev Motwani, and Terry Winograd.
\newblock The pagerank citation ranking: Bringing order to the web.
\newblock Technical report, Stanford InfoLab, 1999.

\bibitem{textrank}
Rada Mihalcea and Paul Tarau.
\newblock Textrank: Bringing order into text.
\newblock In {\em Proceedings of the 2004 conference on empirical methods in
  natural language processing}, pages 404--411, 2004.

\bibitem{Xu2009Detecting}
Wei Xu, Ling Huang, Armando Fox, David Patterson, and Michael~I. Jordan.
\newblock Detecting large-scale system problems by mining console logs.
\newblock In {\em ACM Sigops Symposium on Operating Systems Principles}, pages
  117--132, 2009.

\bibitem{lin2016log}
Qingwei Lin, Hongyu Zhang, Jian-Guang Lou, Yu~Zhang, and Xuewei Chen.
\newblock Log clustering based problem identification for online service
  systems.
\newblock In {\em Proceedings of the 38th International Conference on Software
  Engineering Companion (ICSE)}, pages 102--111. ACM, 2016.

\bibitem{Stanovsky2016oieBenchmark}
Gabriel Stanovsky and Ido Dagan.
\newblock Creating a large benchmark for open information extraction.
\newblock In {\em EMNLP}, 2016.

\bibitem{HeLuheng2015QuestionAnswerDS}
Luheng He, Mike Lewis, and Luke Zettlemoyer.
\newblock Question-answer driven semantic role labeling: Using natural language
  to annotate natural language.
\newblock In {\em EMNLP}, 2015.

\bibitem{Corro2013ClausIECO}
Luciano~Del Corro and Rainer Gemulla.
\newblock Clausie: clause-based open information extraction.
\newblock In {\em WWW '13}, 2013.

\bibitem{Mausam2012OpenLL}
Mausam, Michael Schmitz, Stephen Soderland, Robert Bart, and Oren Etzioni.
\newblock Open language learning for information extraction.
\newblock In {\em EMNLP-CoNLL}, 2012.

\bibitem{Stanovsky2016GettingMO}
Gabriel Stanovsky, Jessica Ficler, Ido Dagan, and Yoav Goldberg.
\newblock Getting more out of syntax with props.
\newblock {\em ArXiv}, abs/1603.01648, 2016.

\bibitem{Angeli2015LeveragingLS}
Gabor Angeli, Melvin Jose~Johnson Premkumar, and Christopher~D. Manning.
\newblock Leveraging linguistic structure for open domain information
  extraction.
\newblock In {\em ACL}, 2015.

\bibitem{rouge}
Chin-Yew Lin.
\newblock {ROUGE}: A package for automatic evaluation of summaries.
\newblock In {\em Text Summarization Branches Out}, pages 74--81, Barcelona,
  Spain, July 2004. Association for Computational Linguistics.

\bibitem{rouge2}
Kavita Ganesan.
\newblock Rouge 2.0: Updated and improved measures for evaluation of
  summarization tasks.
\newblock {\em arXiv preprint arXiv:1803.01937}, 2018.

\bibitem{tf-idf}
Sungjick Lee and Han-joon Kim.
\newblock News keyword extraction for topic tracking.
\newblock In {\em 2008 Fourth International Conference on Networked Computing
  and Advanced Information Management}, volume~2, pages 554--559. IEEE, 2008.

\bibitem{shang2014understanding}
Weiyi Shang, Meiyappan Nagappan, Ahmed~E Hassan, and Zhen~Ming Jiang.
\newblock Understanding log lines using development knowledge.
\newblock In {\em 2014 IEEE International Conference on Software Maintenance
  and Evolution}, pages 21--30. IEEE, 2014.

\bibitem{he2018identifying}
Shilin He, Qingwei Lin, et~al.
\newblock Identifying impactful service system problems via log analysis.
\newblock In {\em Proceedings of the 2018 26th ACM Joint Meeting on European
  Software Engineering Conference and Symposium on the Foundations of Software
  Engineering}, pages 60--70. ACM, 2018.

\bibitem{Blei2003Latent}
David~M Blei, Andrew~Y Ng, and Michael~I Jordan.
\newblock Latent dirichlet allocation.
\newblock {\em Journal of machine Learning research}, 3(Jan):993--1022, 2003.

\end{thebibliography}
